\documentstyle[titlepage]{article}
\def\openone{\leavevmode\hbox{\small1\kern-3.3pt\normalsize1}}

 \def\thefootnote{\fnsymbol{footnote}}
 \def\@makefnmark{\mbox{$^\@thefnmark$}}
\def\thefootnote{\mbox{\noindent$\fnsymbol{footnote}$}}
    \long\def\@makefntext#1{\noindent$^{\@thefnmark}$#1}

\newcommand{\prepr}[3]{\large
\begin{center}
\hspace*{7cm} Preprint
\hspace*{7cm} IFUNAM \, FT-94-{#1}
\hspace*{7cm} EFUAZ\qquad  94-{#2}
\hspace*{7cm}  {#3}\,\,\,1994
\end{center}}

\newcommand{\npb}[2]{{\em Nucl.\ Phys.\ B\ }{\bf #1}{(#2)}}
\newcommand{\pr}[2]{{\em Phys.\ Rev.\ }{\bf #1}{(#2)}}
\newcommand{\prd}[2]{{\em Phys.\ Rev.\  D\ }{\bf #1}{(#2)}}
\newcommand{\prl}[2]{{\em Phys.\ Rev.\ Lett.\ }{\bf #1}{(#2)}}
\newcommand{\pl}[2]{{\em Phys.\ Lett.\ }{\bf #1}{(#2)}}
\newcommand{\pla}[2]{{\em Phys.\ Lett.\ A\ }{\bf #1}{(#2)}}
\newcommand{\plb}[2]{{\em Phys.\ Lett.\ B\ }{\bf #1}{(#2)}}
\newcommand{\epl}[2]{{\em Europhys.\ Lett.\ }{\bf #1}{(#2)}}
\newcommand{\jmp}[2]{{\em J.\ Math.\ Phys.\ }{\bf #1}{(#2)}}

\newcommand{\cjp}[2]{{\em Czech.\ J.\  Phys.\  B\ }{\bf #1}{(#2)}}

\newcommand{\ptp}[2]{{\em Progr.\ Theor.\ Phys.\ }{\bf #1}{(#2)}}

\newcommand{\prsl}[2]{{\em Proc.\  Roy.\  Soc.\ (London)\  A\ }{\bf #1}{(#2)}}
\newcommand{\pps}[2]{{\em Proc.\  Phys.\  Soc.\  }{\bf #1}{(#2)}}
\newcommand{\mpl}[2]{{\em Mod.\ Phys.\ Lett.\ }{\bf #1}{(#2)}}
\newcommand{\ijmp}[2]{{\em Int.\ J.\ Mod.\ Phys.\ }{\bf #1}{(#2)}}
\newcommand{\rmp}[2]{{\em Rev.\ Mod.\ Phys.\ }{\bf #1}{(#2)}}
\newcommand{\apny}[2]{{\em Ann.\ Phys.\ (N.Y.)\ }{\bf #1}{(#2)}}
\newcommand{\jpa}[2]{{\em J. \ Phys. A\ }{\bf #1}{(#2)}}

\newcommand{\jp}[2]{{\em J. \ Phys.\  }{\bf #1}{(#2)}}

\newcommand{\nc}[2]{{\em Nuovo\ Cim. \ }{\bf #1}{(#2)}}
\newcommand{\ncs}[2]{{\em Nuovo\ Cim.\ Suppl. \ }{\bf #1}{(#2)}}
\newcommand{\lnc}[2]{{\em Lett.\  Nuovo \ Cim.\ } {\bf #1}{(#2)}}
\newcommand{\fp}[2]{{\em Found.\  Phys.\ } {\bf #1}{(#2)}}

\newcommand{\rmf}[2]{{\em  Rev. Mex. \ Fiz. \ }{\bf #1}{(#2)}}

\newcommand{\pzh}[2]{{\em  Pis'ma\ ZhETF\  }{\bf #1}{(#2)}}
\newcommand{\zh}[2]{{\em  ZhETF\  }{\bf #1}{(#2)}}
\newcommand{\jl}[2]{{\em JETP\  Lett.\ }{\bf #1}{(#2)}}
\newcommand{\jetp}[2]{{\em JETP\  }{\bf #1}{(#2)}}

\newcommand{\hj}[2]{{\em  Hadronic \ J. \ }{\bf #1}{(#2)}}
\newcommand{\hjs}[2]{{\em  Hadronic \ J. \  Suppl.\ }{\bf #1}{(#2)}}

\newcommand{\zp}[2]{{\em  Zeitshr. \  Phys. \ }{\bf #1}{(#2)}}

\def\theequation{\arabic{section}.\arabic{equation}}
\def\appendix{
\par
\setcounter{equation}{0}
\def\theequation{A1.\arabic{equation}}}

\begin{document}
\thispagestyle{empty}

\title{\vspace*{-3cm}\prepr{44}{01}{March}\vspace*{2cm}
{\large {\bf  NOTES ON OSCILLATOR-LIKE INTERACTIONS OF
VARIOUS SPIN RELATIVISTIC PARTICLES}}\thanks{Talk contributed at  the Second
Workshop  "Osciladores Arm\'onicos".  Cocoyoc, M\'exico, March 23-25,
1994. To be published in the NASA Conference Proceedings.}}\vskip6mm

\author{
{\bf  Valeri V. Dvoeglazov}\thanks{On leave of absence from \,
{\it Dept. Theor. \& Nucl. Phys., Saratov State University, Astrakhanskaya
str., 83, Saratov RUSSIA}}\,\,\,\thanks{Email: {\it valeri@bufa.reduaz.mx,
dvoeglazov@main1.jinr.dubna.su}} \\ \vspace{1mm}
{\it Escuela de F\'{\i}sica, Universidad Aut\'onoma
de Zacatecas}\\
{\it Ciudad Universitaria, Antonio Doval\'{\i} Jaime\, s/n}\\
{\it  Zacatecas, ZAC., M\'exico\,  98068}  \\
and   \\
{\bf Antonio del Sol Mesa}\thanks{Email: {\it
antonio@sysul1.ifisicacu.unam.mx}}\\  \vspace{1mm}
{\it Deptartamento de F\'{\i}sica Te\'orica, Instituto de F\'{\i}sica}\\
{\it Universidad Nacional Aut\'onoma de M\'exico}\\
{\it Apartado Postal 20-364, D. F., M\'exico \, 01000}}

\begin{abstract}
{\large The equations for various spin particles with oscillator-like
interactions are
discussed in this
talk. Contents: 1.  Comment on "The Klein-Gordon Oscillator"; 2. The Dirac
oscillator in quaternion form; 3. The Dirac-Dowker  oscillator; 4. The Weinberg
oscillator; 5. Note on the two-body Dirac oscillator.}
\end{abstract}
\maketitle

\newpage
\renewcommand{\thefootnote}{\alph{footnote}}
\vspace*{1cm}

\section{Comment on "The Klein-Gordon Oscillator"\vskip1mm \normalsize{{\bf
[By S. Bruce and P. Minning,
{\em Nuovo~Cim.} {\bf 106A}  (1993) 711]}}}

Concept of relativistic harmonic oscillator with intrinsic spin structure has
been proposed long ago~\cite{Cook}. However,  in connection with the
publications of Moshinsky {\it et al.}~\cite{Mosh} the interest in this simple
model with  a  $j=1/2$ Hamiltonian that is linear
in both momenta and coordinates has grown recently~\cite{oth}.  Analogous type
of interaction
has been considered for the case of $j=0$ and $j=1$ Duffin-Kemmer
field~\cite{DKP} and for the case of $j=0$ Klein-Gordon field~\cite{KG}.

In the paper~\cite{KG} the operators $\vec Q$, coordinate, and $\vec P$,
momentum, have been
represented in $n \otimes n$ matrix form
\begin{equation}
\vec Q=\hat \eta \vec q, \hspace*{1cm} \vec P=\hat \eta \vec p,
\end{equation}
with $\hat \eta^2=1$. The interaction in the Klein-Gordon equation has been
introduced in
the following way:
\begin{equation}
\vec P\rightarrow\vec P-im\hat\gamma\hat\Omega\cdot\vec Q,
\end{equation}
where for the sake of completeness  $\hat \Omega$ is chosen by $3\otimes 3$
matrix
with coefficients $\hat\Omega_{ij}=\omega_{i}\delta_{ij}$ (the  case of
anisotropic oscillator).
The $\hat\gamma$ matrix obeys the following anticommutation relations:
\begin{equation}
\left \{\hat\gamma,\hat\eta\right \}=0, \hspace*{1cm} \hat\gamma^2=1.
\end{equation}
The Klein-Gordon equation for $\Psi (\vec q, t)$, the wave function which could
be expanded in two-component form, is then
\begin{equation}
-\frac{\partial^2}{\partial t^2}\Psi (\vec q, t)= \left (\vec p^{\,2}+m^2\vec
q\cdot\hat\Omega^2\cdot\vec q+m\hat\gamma\,\, tr\Omega+m^2\right )\Psi (\vec q,
t),
\end{equation}
what gives the energy spectrum~\cite{KG}
\begin{eqnarray}
E^2_{(a)N_i}-m^2&=&2m\left (\omega_1 N_1+\omega_2 N_2+\omega_3 N_3\right
),\hspace*{1cm}
N_1, N_2, N_3=0, 1, 2 \ldots\nonumber\\
E^2_{(b)N_i}-m^2&=&2m \left  (\omega_1 (N_1+1)+\omega_2 (N_2+1)+\omega_3
(N_3+1)\right ).
\end{eqnarray}
It  becomes, up to an additive constant,  the  spectrum of the anisotropic
oscillator  in the non-relativistic limit.

However, the physical sense of implementing the matrices $\hat\eta$ and
$\hat\gamma$ in~\cite{KG}  is
obscure. In this Section we try to attach some physical  foundations to this
procedure.

It is well-known some ways to recast the Klein-Gordon equation in the
Hamiltonian form~\cite{st}-\cite{FV}. First of all, the Klein-Gordon equation
could be re-written to the system
of two coupled equations [8,p.98]
\begin{equation}
\frac{\partial\Psi}{\partial x^{\,\alpha}}=\kappa \Xi_{\,\alpha}, \hspace*{1cm}
\frac{\partial \Xi^{\,\alpha}}{\partial x^{\,\alpha}}=-\kappa\Psi,
\end{equation}
where  $\kappa=mc/\hbar$ (in the following we use the system where
$c=\hbar=1$).
By means of redefining the components
they are easy to present in the matrix Hamiltonian form (cf. with~\cite{KG-d})
\begin{equation}\label{eq:hami}
i\frac{\partial}{\partial t} \left (\matrix{
\phi\cr
\chi_1\cr
\chi_2\cr
\chi_3\cr
}\right )=\left [\left (\matrix{
0 & p_1 & p_2 & p_3 \cr
p_1 & 0 & 0 & 0 \cr
p_2 & 0 & 0 & 0 \cr
p_3 & 0 & 0 & 0 \cr
}\right ) +m\left (\matrix{
1 & 0 & 0 & 0 \cr
0 & -1 & 0 & 0 \cr
0 & 0 & -1 & 0 \cr
0 & 0 & 0 & -1 \cr
}\right )\right ]\left (\matrix{
\phi\cr
\chi_1\cr
\chi_2\cr
\chi_3\cr
}\right )
\end{equation}
provided that
\begin{eqnarray}
\cases{\phi=i\partial_t \Psi+m\Psi &\cr
\chi_i=-i \bigtriangledown_i \Psi = \vec p_i \Psi .& }
\end{eqnarray}
Using matrices
\begin{equation}
\vec \alpha=\pmatrix{
0_{1\otimes 1} & \mid & {\bf i} & {\bf  j} & {\bf k} \cr
\hline
{\bf i}  & \vert & & & \cr
{\bf  j}  & \vert &  &0_{3\otimes 3}& \cr
{\bf k} & \vert &  & &
},\hspace*{1cm} \beta=\pmatrix{
\openone_{1\otimes 1} & \mid & 0 \cr
\hline
&\vert & \cr
0 & \vert & -\openone_{3\otimes 3} \cr
& \vert & \cr
}
\end{equation}
(${\bf i}, {\bf j}, {\bf k}$ are the orth-vectors of  the Euclidean basis)
and substituting analogously [2a], i. e. $\vec p\rightarrow \vec
p-im\omega\beta\vec r$,
we come to the equation for upper component
\begin{equation}\label{eq:kgo}
(E^2-m^2)\phi=\left [\vec p^{\,2}+m^2\omega^2\vec r^{\,2}-3m\omega\right ]\phi
\end{equation}
what coincides with Eq. (10a) of ref.~\cite{KG} in the case
$\omega_1=\omega_2=\omega_3$.

The similar formulation also originated from the Duffin-Kemmer
approach~\cite{DKP2}.
In this case the wave function  $\Phi=column (\phi_1, \phi_2, \chi_i)$  is
five-dimensional and its components are
\begin{eqnarray}
\cases{\phi_1=(i\partial_t \Psi + m\Psi)/\sqrt{2} & $$\cr
\phi_2=(i\partial_t \Psi - m\Psi)/\sqrt{2} & $$\cr
\chi_i=-i \bigtriangledown_i \Psi = \vec p_i \Psi. & $$}
\end{eqnarray}
It satisfies the equation
\begin{equation}\label{eq:dfk}
i\frac{\partial \Phi}{\partial t}= \left (\vec B \vec p +m\beta_0\right )
\Phi,\hspace*{1cm} B_\mu= [\beta_0, \beta_\mu]_{-}
\end{equation}
(our choice of $5\otimes 5$ dimension $\beta$-matrices corresponds to ref.
[4b]). As shown there, the substitution $\vec p\rightarrow \vec p - im\omega
\eta_0 \vec r$, where
\begin{eqnarray}
\eta=\pmatrix{
\openone_{2\otimes 2} &\mid & 0 \cr
\hline
& \vert &\cr
0 & \vert & - \openone_{3\otimes 3}\cr
& \vert &\cr
},
\end{eqnarray}
leads to the equation (\ref{eq:kgo}) for both $\phi_1$ and $\phi_2$. Let us
remark, in both the approach based on Eq. (\ref{eq:hami}) and the Duffin-Kemmer
 approach, Eq. (\ref{eq:dfk}), we have the equation
\begin{equation}
(E^2 -m^2)\chi_i=(p_i -im\omega x_i) (p_j +im\omega x_j)\chi_j
\end{equation}
for down component,
which seems to  not  be reduced  to oscillator-like equation.

Then, Sakata-Taketani approach~\cite{ST,FV} is characterized by  the equation
which we write in the form:
\begin{equation}
i {\partial \over \partial t}\Phi=\left \{\frac{\vec p (\tau_3+i\tau_2) \vec
p}{2m}+m\tau_3\right \} \Phi,
\end{equation}
with $\tau_i$  being the Pauli matrices. $\Phi=column (\phi, \chi)$ is the
two-component wave function with components which could be written as
following:
\begin{eqnarray}
\cases{\phi=(\Psi +{i\over m}\partial_t \Psi)/\sqrt{2}& $$\cr
\chi=(\Psi - {i\over m}\partial_t \Psi)/\sqrt{2}.& $$}
\end{eqnarray}

{}From the previous experience we learned that in order to get the
oscillator-like equation we need to do substitution with matrix which
anticommutes with matrix structure of the momentum part of the equation. In our
case the matrix  which has this property  is $\tau_1$ matrix.  Therefore, we do
the substitution
\begin{equation}\label{eq:subs}
\vec p \rightarrow \vec p -im\omega\tau_1 \vec r \nonumber
\end{equation}
and come to
\begin{equation}
E^2 \xi=\left [\vec p^{\,2}+m^2\omega^2\vec r^{\,2}-3m\omega+m^2\right ]\xi,
\end{equation}
where $\xi=\phi+\chi$
and to the analogous equation for $\eta=\phi-\chi={E\over m}(\phi+\chi)$.  In
the process of calculations we  convinced ourselves that the interaction
Hamiltonian
\begin{equation}
{\cal H}={1\over 2m}\left (\vec p^{\,2}+m^2 \omega^2 \vec
x^{\,2}-3m\omega\right ) (\tau_3 +i\tau_2) +m \tau_3
\end{equation}
is the same as in [4b,formula (3.9)] since $\tau_1 (\tau_3 +i \tau_2) \tau_1 =
- (\tau_3 +i\tau_2)$ and $(\tau_3+i\tau_2)\tau_1=\tau_3 + i\tau_2$.

Let us also recall the  Dirac oscillator in $(1+1)$ dimensions [3j].
Curiously, the formula  (\ref{eq:hami}) in $(1+1)$ dimensions  looks like  the
Dirac equation in the case of the choice
of $\gamma$-matrices as in [3j], i. e.
\begin{equation}
\alpha= \tau_x \hspace*{1cm} \beta=\tau_z\nonumber
\end{equation}
In the case of the choice of $\gamma$-matrices as in~\cite{Cooper}, i.~e.
\begin{equation}
\gamma_0=\tau_1 \hspace*{1cm} \gamma_1=i\tau_3,
\end{equation}
in order to obtain the Dirac oscillator it is necessary to do substitution
(\ref{eq:subs}).
\vspace*{1cm}

\setcounter{equation}{0}
\section{The Dirac oscillator in quaternion form}

The quaternion (and conjugated to it) with real coefficient is defined as
\begin{eqnarray}
q&=&q_0+iq_1+jq_2+kq_3\nonumber\\
\bar q&=&q_0-iq_1-jq_2-kq_3.
\end{eqnarray}
The basis vectors satisfy the equations $i^2=j^2=k^2=-1$ and $ij=-ji=k$ with
cyclic permutations.

Considering a two-component quaternionic spinor (or $SL(2,H)$ spinor) one could
write
the free Dirac equation as following, ref. [14c,d]\footnote{Let us mention
other
proposed equations, e. g.~\cite{Conway}-\cite{pr90}.  However, the equation
given by J. Sou\v{c}ek~\cite{sou}
\begin{equation}
(i\partial_1 +j\partial_2 +k\partial_3)\psi=(-\sqrt{-1}\tau_2 \partial_0
+\tau_3 m)\psi
\end{equation}
leads to the tachionic solutions of the  Dirac equation. For a gauge theory,
involving quaternion-valued fields, see~\cite{Ch}-\cite{HJ}.  Meantime, in the
biquaternion formulation of Morita~\cite{Morita} the norm  $N(q)=\bar q q =
q\bar q$ is not well defined. However, in the case when  components of
quaternionic spinor are the grassmanian numbers,  the norm vanishes:
\begin{equation}
N(q)=  -\eta_{ij} q^i q^j \equiv (q_0)^2 - (q_1)^2 - (q_2)^2 - (q_3)^2=0.
\end{equation}
See {\em footnote} on p. 1651 of ref. [14a] where the necessity of this
suggestion is discussed.
}
\begin{equation} \label{eq:DE}
\tilde \Gamma\cdot \partial {\bf\Psi} - m\tau_3 {\bf\Psi} k=0.
\end{equation}
Anticommutation relations for $\tilde\Gamma$ are given in [14d,p.222].
In Pauli representation ($i\rightarrow \, -\sqrt{-1} \tau_1$,  $j\rightarrow \,
-\sqrt{-1} \tau_2$ and
$k\rightarrow \, -\sqrt{-1} \tau_3$) it goes through to usual Dirac equation
and its complex conjugate.
As mentioned in~\cite{Morita} it is convenient to diagonalize the matrices
entering in Eq. (\ref{eq:DE}) using matrix
\begin{eqnarray}
T={1\over \sqrt{2}}\pmatrix{
1 & -\sqrt{-1} \cr
1 & \sqrt{-1} \cr
}.
\end{eqnarray}
In such a way we come to biquaternionic formulation ($q_i \in {\bf C}$, i.~e.
the coefficients are the complex numbers):
\begin{eqnarray}
&&\bar\partial \psi_L +im\bar \psi^\dagger_R=0\label{eq:e1}\\
&&\partial\bar\psi_R^\dagger +im \psi_L=0,\label{eq:e2}
\end{eqnarray}
where $\psi_L \equiv \psi p_+$, $\psi_R \equiv \psi p_{-}$. This decomposition
of ${\bf \Psi}$ into
left ideals is carried out by means of the projection operators $p_{\pm}=(b_0
\pm b_3)/2$.  New basis is $b_0\equiv 1, b_{1}\equiv \sqrt{-1} i,  b_{2}\equiv
\sqrt{-1} j, b_{3}\equiv \sqrt{-1} k$ and $\bar b_0 =b_0, \bar b_\alpha
=-b_\alpha$.
Introducing interaction in the form $\partial_i \rightarrow \partial_i + \tau_3
V_i(\vec x)$, V is the compensating field for this type of $Sp (1,Q)$
transformations, and taking into account that the vectors of biquaternionic
basis anticommute $b_\alpha \bar b_\beta +b_\beta \bar b_\alpha =
-2\eta_{\alpha\beta}$, $\eta_{\alpha\beta}= diag (-1, 1, 1, 1)$,  we come
to the equations for the  left and right  spinor-quaternions in the following
form:
\begin{eqnarray}
(E^2 - m^2)\psi_L &=& \left [ (\vec p^{\,2}+ k^2 \vec x^{\,2}) - 3k -
2\epsilon_{ijk} b_k  x_i p_j\right ]\psi_L \label{eq:1}\\
(E^2 - m^2)\bar\psi_R^\dagger &=& \left [ (\vec p^{\,2}+ k^2 \vec x^{\,2}) + 3k
 +2\epsilon_{ijk} b_k  x_i p_j\right ]\bar\psi_R^\dagger\label{eq:2}
\end{eqnarray}
if we  choose $V_i(\vec x)=k x_i$. Eqs. (\ref{eq:1}) and (\ref{eq:2}) are  the
Dirac oscillator equations in the Pauli rep, $b_k \rightarrow \tau_k$.
Analogous equations for $\psi_R$ and $\bar \psi_L^{\dagger}$ could be obtained
if we start from  (\ref{eq:e1}) and (\ref{eq:e2}) with the opposite signs at
the mass terms.
In the above,  we assume, following for Morita~\cite{Morita}, that the
imaginary unit $\sqrt{-1}$ commutes with Hamilton's basis vectors $i, j, k$,
what is not obviously.  Consideration of $\sqrt{-1}$  on  an equal footing with
the Hamilton's units would also be interesting.

The investigation of interactions of quaternionic Dirac field deserves further
elaboration.
\vspace*{1cm}

\setcounter{equation}{0}
\section{The Dirac-Dowker  oscillator}

In this Section we start from the equation for any spin given by
Dirac~\cite{Dirac}, see also~\cite{Fiertz}, in the form
written down by Corson, ref. [8,p.154],  (here we  use Corson's notation)
\begin{eqnarray}\label{eq:DD}
\cases{\partial^{\dot A B}v_B (k+{1\over 2}) \psi (k+{1\over 2}, l-{1\over
2})-m \left (\frac{2k+1}{2l}\right )^{1/2}v^{\dot A}(l) \psi (k,l)=0&\cr
&\cr
\partial_{\dot A B}v^{\dot A} (l) \psi (k, l)+m \left (\frac{2l}{2k+1}\right
)^{1/2} v_{B} (k+{1\over 2}) \psi (k+{1\over 2},l-{1\over 2})=0&}
\end{eqnarray}
where $v_A$ and $v_{\dot A}$ are the rectangular spinor-matrices of  $2k$ rows
and $2k+1$ columns (see, e. g., section 17b of ref.~\cite{b-st1} for details).
The wave function $\psi (k,l)$ belongs to the $(k,l)$ representation of the
homogeneous Lorentz group.
The choice $l=1/2$ and $k=j-1/2$, $j$ is the spin of a particle, permits one to
reduce a number of
subsidiary conditions. Moreover, the equations (\ref{eq:DD}) are shown  by
Dowker~\cite{Dowker} to recast to the
matrix form which is similar to the well-known Dirac equation for $j=1/2$
particle
\begin{eqnarray}\label{eq:DDo}
\alpha^\mu \partial_\mu \Phi &=& m\Upsilon ,\nonumber\\
\bar\alpha^\mu \partial_\mu \Upsilon &=& -m \Phi .
\end{eqnarray}
The $4j$- component function $\Phi$ could be identified with the wave function
in $(j,0)\oplus (j-1,0)$ representation. Then,  $\Upsilon$, which also has $4j$
components, is written down
\begin{eqnarray}
\Upsilon = (-1)^{2j} (2j)^{-{1\over 2}}\left (\matrix{
v_{\dot A} (j-{1\over 2})\otimes v^{\dot A} ({1\over 2})\cr
u_{\dot A} (j) \otimes v^{\dot A} ({1\over 2})\cr
}\right ) \psi (j-{1\over 2}, {1\over 2}).
\end{eqnarray}
and it belongs to $(j-1/2,1/2)$ representation.
The matrices $\alpha^\mu$ and $\bar\alpha^\mu=\alpha_\mu$  obeys the
anticommutation
relations of Pauli matrices
\begin{equation}\label{eq:ac}
\bar \alpha^{(\mu} \alpha^{\nu)}=g^{\mu\nu} .
\end{equation}
This set of matrices has been investigated in details in ref. [24b,c] and
$\alpha^\mu$  was proved there to satisfy all the algebraic relations of the
Pauli matrices except for completeness.

Defining $p_\mu=-i\partial_\mu$ and the analogs of $\gamma$-  matrices as
following:
\begin{eqnarray}\label{eq:ga}
\gamma^\mu = \pmatrix{
0 & -i\bar\alpha^\mu \cr
i\alpha^\mu & 0 \cr
}
\end{eqnarray}
the set of equations (\ref{eq:DDo}) is written down to the form of the Dirac
equation
\begin{equation}\label{eq:Di}
\left (p_\mu \gamma^\mu - m\right ) \left (\matrix{
\Phi\cr
\Upsilon\cr
}\right ) =0.
\end{equation}
However, let us not forget that $\Phi$ and $\Upsilon$ are 2-spinors only in the
case of $j=1/2$.

As mentioned in, e. g., ref. [8,p.33,124], in the case of spin $j=1/2$ the set
of $\gamma$- matrices in representation (\ref{eq:ga})
\begin{eqnarray}
\gamma^0 &=& \pmatrix{
0 & -i \openone_{2\otimes 2}\cr
i \openone_{2\otimes 2} & 0 \cr
}, \hspace*{1cm} \gamma^1=\pmatrix{
0 & i\tau^1 \cr
i\tau^1 & 0 \cr
}\nonumber\\
& &\nonumber \\
\gamma^2 &=& \pmatrix{
0 & i\tau^2  \cr
i \tau^2 & 0 \cr
}, \hspace*{1cm} \gamma^3=\pmatrix{
0 & i\tau^3 \cr
i\tau^3 & 0 \cr
}
\end{eqnarray}
is defined up to the unitary transformation and  Eq. (\ref{eq:Di}) could be
recast to the Hamiltonian form given by Dirac (with $\alpha_k$ and $\beta$
matrices) by means of the unitary matrix. It is easy to carry out the same
procedure ($\alpha^k= {\cal S} \gamma^0\gamma^k {\cal S}^{-1}$ and $\beta={\cal
S} \gamma^0 {\cal S}^{-1}$) for $\gamma$ matrices (Eq. \ref{eq:ga}) and
functions of arbitrary spin ($\Psi= {\cal S}^{-1} \Phi$).  For our aims it is
convenient to chose the unitary matrix as following:
\begin{eqnarray}
{\cal S} ={1\over \sqrt{2}}\pmatrix{
\openone_{4j\otimes 4j} & i \openone_{4j\otimes 4j} \cr
i \openone_{4j\otimes 4j} & \openone_{4j\otimes 4j} \cr
}.
\end{eqnarray}
 After standard substitution $\vec p \rightarrow \vec p -im\omega \gamma^0 \vec
r$ we obtain
\begin{eqnarray}
\cases{E\phi = -i\left [ \alpha_0 (\vec \alpha\vec p) +im\omega (\vec \alpha
\vec r)\right ]\upsilon +m\alpha_0\phi, & $$\cr
E\upsilon = i \left [ \alpha_0(\vec \alpha \vec p) -im\omega (\vec \alpha \vec
r)\right ]\phi -m \alpha_0\upsilon. & $$}
\end{eqnarray}
Since it follows from the anticommutation relations (\ref{eq:ac}) that
$\alpha_i \alpha_0=\alpha_0 \alpha_i$ we have the equations
which coincide with Eq. (8)  of ref. [2a] or  Eqs. (3.6) and (3.12) of ref.
[2h] except for $ \tau_\mu \rightarrow  \alpha_\mu$, i.~e. their explicit
forms,
\begin{eqnarray}
(E^2 -m ^2)\phi &=& \left [ \vec p^{\,2}+ m\omega \vec r^{\,2} -3\alpha_0
m\omega -  m\omega \alpha_0 \alpha^{ [ i}\bar\alpha^{j ]} r^i
\bigtriangledown_j\right ]\phi\\
(E^2 +m ^2)\upsilon &=& \left [\vec p^{\,2}+ m\omega \vec r^{\,2} +3\alpha_0
m\omega +  m\omega \alpha_0 \alpha^{[ i}\bar\alpha^{j ]} r^i
\bigtriangledown_j \right ]\upsilon.
\end{eqnarray}
Thus,  we  convinced ourselves that we got the same oscillator-like interaction
and the similar spectrum as for the case of $j=1/2$ particles in  [2a].
 \vspace*{1cm}

\setcounter{equation}{0}
\section{The Weinberg oscillator}

The principal equation of  $2(2j+1)$- component approach~\cite{Weinberg} in the
case of spin $j=1$  is
\begin{equation}
(\gamma_{\mu\nu}p_\mu p_\nu + M^2)\Psi^{(j=1)}(x)=0,
\end{equation}
with $\gamma_{\alpha\beta}$ being defined by the formulae
\begin{equation}\label{eq:gam}
\gamma_{ij} \equiv\pmatrix{
0 & \delta_{ij}-S_i S_j- S_j S_i \cr
\delta_{ij}-S_i S_j- S_j S_i & 0 \cr
},
\end{equation}
\begin{equation}
\gamma_{i4}=\gamma_{4i} \equiv\pmatrix{
0 & iS_i \cr
-iS_i & 0 \cr
},\ \ \ \ \ \gamma_{44} \equiv\pmatrix{
0 & \openone_{3\otimes 3} \cr
\openone_{3\otimes 3} & 0\cr
};
\end{equation}
($S_i$ are the  spin matrices for a vector particle).

The $j=1$  Hamiltonian  has been  given in refs.~\cite{Weaver,Mathews}:
\begin{equation}
{\cal H}=\frac{2E^2}{2E^2-M^2}(\vec\alpha\vec p)+\beta \left
[E-\frac{2E}{2E^2-M^2}(\vec\alpha\vec p)^2\right ],
\end{equation}
where
\begin{center}
$\vec\alpha\equiv\pmatrix{
\vec S & 0 \cr
0 & -\vec S \cr
}$,\hspace*{5mm}
$\beta\equiv\pmatrix{
0 & \openone_{3\otimes 3} \cr
\openone_{3\otimes 3} & 0\cr
}$.\\
\end{center}

Though this way of description is a little bit antique,  attention
has again been paid to it recently, e. g.~\cite{Ahlu,Dvoeglaz}. This formalism
presents oneself
the example of the Bargmann-Wightman-Wigner type quantum field theory [28b].
The remarkable feature is the fact that boson and its antiboson have the
opposite
parities.

In general, the upper and down components of  6- component wave function do not
uncouple neither under the interaction  $\vec p \rightarrow \vec p -im\omega
\beta\vec r$ nor under $\gamma_{5,\mu\nu} u_\mu r_\nu$. However, if
we introduce the Dirac oscillator interaction so
 that the conditions of the longitudity of $\Psi =column (\phi_i, \chi_i)$
respective to $\vec r$
\begin{equation}
{\bf \vec r} {\bf \times} \vec {\bf \phi}  =  0, \quad {\bf \vec r} {\bf
\times} \vec {\bf \chi} = 0
\end{equation}
are fulfilled,
we come to more simple equations ($\xi=\phi - \chi$, $\eta= \phi +\chi$)
\begin{eqnarray}
(2E^2 - M^2) \xi &=&  E(\vec S \vec p)\eta +\left [ (\vec S\vec p) -
k (\vec S \vec r)\right ] (\vec S \vec p)\xi ,\\
E(\vec S \vec p) \xi &=& \left [ (\vec S \vec p) + k (\vec S \vec r)\right ]
(\vec S \vec p) \eta
\end{eqnarray}
which could be uncoupled to the following form ($k=im\omega$)
\begin{eqnarray}
(\vec S \vec p) (E^2 - M^2) (\vec S \vec p) \xi &=& (\vec S \vec p)
\left [ \vec p^{\,2} + m\omega^2 \vec r^{\,2} +3m\omega +4m\omega
\vec S [\vec r \times \vec p]\right ] (\vec S \vec p) \xi\nonumber\\
&&\\
(\vec S \vec p) (E^2 - M^2) (\vec S \vec p) \eta &=& (\vec S \vec p)
\left [ \vec p^{\,2} + m\omega^2 \vec r^{\,2} -3m\omega -4m\omega
\vec S [\vec r \times \vec p]\right ] (\vec S \vec p) \eta-\nonumber\\
&&\qquad -im\omega (2E^2 - M^2) (\vec S \vec r) (\vec S \vec p) \eta
\end{eqnarray}

These  equations can be considered as the extension of the equations with Dirac
oscillator interaction  to the $j=1$ case, for the
components  $(\vec S\vec p)\xi$ and  $(\vec S\vec p)\eta$.  However, remark
that one has
the additional spin-orbit term acting as earlier  at  $\eta$.

\vspace*{1cm}

\setcounter{equation}{0}
\section{Note on the two-body Dirac oscillator}

The two-body Dirac Hamiltonian with oscillator-like interaction is given by
(see, e. g., refs. [2d,f])
\begin{eqnarray}
\lefteqn{i\left ({\partial \over \partial t_1} + {\partial \over \partial
t_2}\right )\psi={\cal H}\psi=}\nonumber\\
&=&\left [{1\over \sqrt{2}}(\vec \alpha_1+\vec \alpha_2)\cdot \vec P+{1\over
\sqrt{2}}(\vec \alpha_1
-\vec \alpha_2)\cdot \vec p-{i\over \sqrt{2}}(\vec \alpha_1-\vec \alpha_2)\cdot
\vec r B+m(\beta_1+\beta_2)\right ] \psi,\nonumber\\
&&
\end{eqnarray}
where the matrices are given by the direct products
\begin{equation}
\vec \alpha_1=\pmatrix{
0 & \vec\sigma_1 \cr
\vec \sigma_1 & 0 \cr
}\otimes \pmatrix{
\openone_{2\otimes 2} & 0 \cr
0 & \openone_{2\otimes 2} \cr
},\quad \vec \alpha_2=\pmatrix{
\openone_{2\otimes 2} & 0 \cr
0 & \openone_{2\otimes 2} \cr
}\otimes \pmatrix{
0 & \vec \sigma_2 \cr
\vec \sigma_2 & 0 \cr
},
\end{equation}
\begin{equation}
B=\beta_1\otimes \beta_2=\pmatrix{
\openone_{2\otimes 2} & 0 \cr
0 & -\openone_{2\otimes 2} \cr
}\otimes \pmatrix{
\openone_{2\otimes 2} & 0 \cr
0 & -\openone_{2\otimes 2} \cr
},
\end{equation}
\begin{equation}
\Gamma_5 = \gamma_1^5\otimes \gamma_2^5 = \pmatrix{
0 & \openone_{2\otimes 2} \cr
\openone_{2\otimes 2} & 0 \cr
}\otimes \pmatrix{
0 & \openone_{2\otimes 2} \cr
\openone_{2\otimes 2} & 0 \cr
}.
\end{equation}
If we are in the center of mass system (c.m.s.) it is possible
to equate $\vec P=0$.

Now we apply the same procedure like that   was used for
transformation the Bargmann-Wigner equation to the Proca
equations (see, e. g., ~\cite[p.30-31]{Lurie}). The 16- component
wave function of two-body Dirac equation could be expanded
on the complete set of matrices:
 $(\gamma^\mu C)$, $(\sigma^{\mu\nu} C)$
and $C$, $(\gamma^5 C)$, and $(\gamma^5 \gamma^\mu C)$.
We consider the system multiplied by $C$, the matrix of
charge conjugation, in order to  trace for symmetric properties
under oscillator-like potentials, see also~\cite[p.31]{Lurie}.
The wave function is decomposed
in symmetric and antisymmetric parts using the above-mentioned
complete system of matrices
\begin{equation}
\psi=\psi_{\{\alpha\beta\} } +\psi_{\left [\alpha\beta\right ]},
\end{equation}
where
\begin{eqnarray}
\psi_{\{\alpha\beta\} }&=& \gamma^\mu_{\alpha\eta} C_{\eta\beta} A_\mu
+\sigma^{\mu\nu}_{\alpha\eta} C_{\eta\beta} F_{\mu\nu}\\
\psi_{\left [\alpha\beta\right ] }&=& C_{\alpha\beta}\phi
+\gamma^5_{\alpha\eta}
C_{\eta\beta} \tilde\phi +\gamma^5_{\alpha\epsilon}\gamma^\mu_{\epsilon\eta}
C_{\eta\beta} \tilde A_\mu.
\end{eqnarray}
In such a way we obtain the set of equations:
\begin{eqnarray}\label{eq:DOP}
&&EA_0=0, \quad E\tilde A_0=-2m\tilde \phi\\
&&E\phi =2i\sqrt{2}  (\vec p_i -i\vec r^{\,i}) F^{i0}\\
&&E\tilde \phi = -2m\tilde A_0 + \sqrt{2}\epsilon_{ijk} (\vec p_i +i\vec
r^{\,i}) F^{jk}\\
\label{eq:58}&&E\tilde A^i = -i\sqrt{2}\epsilon_{ijk} (\vec p_j \mp i\vec
r^{\,j}) A^k  \\
\label{eq:59}&&EA^i = 4im F^{0i} +i\sqrt{2} \epsilon_{ijk} (\vec p_j \pm i\vec
r^{\,j})\tilde A^k\\
&&EF^{0i} = -2im A^i + i\sqrt{2} (\vec p_i +i\vec r^{\,i}) \phi\\
&&E F_{jk} = {1\over \sqrt{2}}\epsilon_{ijk} (\vec p_i -i\vec
r^{\,i})\tilde\phi
\end{eqnarray}
Let us mention that for another type of Dirac oscillator-like interaction $\sim
(\vec \alpha_1 -\vec \alpha_2) B\Gamma_5$  the only changes are the sign
changes at the term  $i\vec r$ in the fourth
and fifth equations of the above system, Eqs. (\ref{eq:58}) and (\ref{eq:59}).
The two-body Dirac oscillator equations in the form (5.8)-(5.14) could be
uncoupled on the set containing  only
functions $\phi$, $\tilde\phi$ and $\tilde A_\mu$ and the another one
containing only $A_\mu$
and $F_{\mu\nu}$:
\begin{eqnarray}
&&(E^2 - 8m^2)\phi = 4(\vec p_i -i\vec r^{\,i})(\vec p_i +i\vec r^{\,i})\phi -
{1\over E} \left \{ {16m \epsilon_{ijk}\vec r^{\,i} \vec p_j \choose 0}\right
\}\tilde A^{\,k}\nonumber\\
&&(E^2 -4m^2)\tilde \phi = 2 (\vec p_i +\vec r^{\,i})(\vec p_i - i\vec
r^{\,i})\tilde \phi\\ \bigskip
&&E\tilde A_0 = -2m\tilde \phi\\ \bigskip
&& (E^2 -8m^2)\tilde A^{\,i} = 2 (\vec p_j \mp i\vec r^{\,j})(\vec p_i \pm
i\vec r^{\,i})\tilde A^{\,j} - 2(\vec p_j \mp i\vec r^{\,j})(\vec p_j \pm i\vec
r^{\,i})\tilde A^{\,i} +\nonumber\\
&&\qquad\qquad+{1\over E} \left \{ {16m\epsilon_{ijk} \vec r^{\,j}\vec p_k
\choose 0}\right \} \phi
\end{eqnarray}
and
\begin{eqnarray}
&&EA_0 = 0\\ \bigskip
&&(E^2 -8m^2) F^{0i} = 4 (\vec p_i +i\vec r^{\,i})(\vec p_j -i\vec r^{\,j})
F^{0j} -\nonumber\\
&&\qquad\qquad-4i{m\over E} (\vec p_j \pm i\vec r^{\,j})(\vec p_i \mp i\vec
r^{\,i}) A^j +4i{m\over E} (\vec p_j \pm i\vec r^{\,j}) (\vec p_j \mp i\vec
r^{\,j}) A^i \nonumber\\
&&E^2 A^i = 2 (\vec p_j \pm i\vec r^{\,j}) (\vec p_i \mp i\vec r^{\,i})A^j -2
(\vec p_j \pm i\vec r^{\,j}) (\vec p _j \mp i\vec r^{\,j})A^i + 4im E
F^{0i}\nonumber\\
&&\\
&& (E^2 -4m^2) F^{jk} = \epsilon_{ijk} \epsilon_{lmn} (\vec p_i -i\vec
r^{\,i})(\vec p_l +i\vec r^{\,l}) F^{mn}.
\end{eqnarray}
This fact proves the Dirac oscillator interaction, like the case of
 introduction of electrodynamic interaction
in the Proca or the Bargmann-Wigner equations,  does not mix $S=1$ and $S=0$
states.

Next, the interaction term of the following form:
\begin{equation}\label{eq:int}
{\cal V}^{int}= {1\over r}\frac{dV(r)/dr}{1-[V(r)]^2} (\vec \alpha_1
-\vec \alpha_2) B\Gamma_5 \vec r
\end{equation}
has been deduced~\cite{Mosh2} from the equation of Relativistic Quantum
Constraint Dynamics (RQCR), e.g. refs.~\cite{sazdjian,crater}, or
$N$- particle Barut equation~\cite{Barut}. In~\cite{Mosh2} it  proved
 to lead to the  Dirac oscillator-like interactions provided that the
definite choice of the
function $V(r)$.
In connection with that let us remark the curious behavior of
the another potential $V(r)$ which has been proposed in ref. [35b,36]:
\begin{equation}\label{eq:quaska}
V(r)= -g^2 \frac{coth\, (rm\pi)}{4\pi r}= -g^2\frac{coth\, (\kappa r)}{4\pi r}.
\end{equation}
It could be deduced from the one-boson exchange quasipotential
$V(\vec p, \vec k;  E)=-g^2 (p-k)^{-2}$ by means of the transformation
into the relativistic configurational representation (RCR) using the
complete set of Shapiro plane-wave functions~\cite{Shapiro}
\footnote{In the quasipotential approach of Kadyshevsky~\cite{kadysh}
it is convenient to pass through to the variables $\Delta_0=
(k_0 p_0 -\vec k \vec p)/m$ and $\vec \Delta=\vec k (-)\vec p=
\vec k-{\vec p\over  m}\left ( k_0 -\frac{\vec k\vec p}{p_0+m}\right )$,
which have the physical sense of the momentum transfer in the Lobachevsky
space, $p_0 -\vec p^{\,2}=m^2$. Then, the transformation to the RCR
is carried out by means of $\xi (\vec \Delta, \vec r)$.}:
\begin{equation}
\xi (\vec p, \vec r)= (p_0 -\vec p \vec n/m)^{-1-irm},\quad p_0=
\sqrt{\vec p^{\,2}+m^2}, \quad\vec n=\vec r/\vert \vec r\vert.
\end{equation}

In the case of the quasipotential (\ref{eq:quaska}) the
interaction term ${\cal V}$, Eq. (\ref{eq:int}), has the different asymptotic
behavior in three regions ($g^2/(4\pi)=1$):
\begin{eqnarray}
\lefteqn{{\cal V}^{int} \simeq {1\over r (r^2-1)} (\vec \alpha_1 -\vec
\alpha_2) B\Gamma_5
\vec r \simeq}\nonumber\\
\medskip
&\simeq&\cases{ (1 /r^3)(\vec \alpha_1 -\vec \alpha_2) B\Gamma_5\vec r,&
if \quad $r>> {1\over \kappa}$\quad  and \quad $r>1$\cr
& $$\cr
-(1/r)  (\vec \alpha_1-\vec \alpha_2) B\Gamma_5\vec r,&if \quad ${1\over
\kappa}<<r<1$,}
\end{eqnarray}behavior
in the infrared region ($r>>{1\over \kappa}$, large distances); and
\begin{equation}
{\cal V}^{int}\simeq -2\kappa (\vec \alpha_1 -\vec \alpha_2) B\Gamma_5\vec
r,\qquad
\mbox{if}\quad  r<<{1\over \kappa},
\end{equation}
in the ultraviolet region (small distances).
We can convince ourselves that in one of the regions we have
the Coulomb-like behavior, in the other region, the Dirac oscillator-like
behavior. This fact could be some quantum field foundations for implementing
the Dirac oscillator potential.

We greatly appreciate valuable discussions with Profs. M.~Moshinsky  and
Yu.~F.~ Smirnov.
Authors also acknowledge very much  Prof.  J. S. Dowker  for sending reprints
of his articles.

\end{document}